\documentclass[submit]{smj}

\usepackage{hyperref}
\usepackage{amsfonts}
\usepackage{amsmath, amsthm, amssymb}
\usepackage{bm}
\usepackage{bbm}
\usepackage[onehalfspacing]{setspace}
\usepackage{pdfpages}
\usepackage{rotating}
\usepackage{lscape}
\usepackage{floatrow}
\usepackage{soulutf8}

\usepackage{color}
\newcommand{\matr}[1]{\bm{#1}}
\newcommand{\bt}{\tilde{\bm{\beta}}}  
\newcommand{\bvec}{\bm{\beta}}  
\newcommand{\kmat}{\bm{K}}

\newcommand{\gvec}{\bm{\gamma}}

\Author{Manuel Carlan\Affil{1}
        and Thomas Kneib\Affil{1}
}
\AuthorRunning{Manuel Carlan and Thomas Kneib}

\Affiliations{

\item Department of Statistics, 
      Faculty of Economics, 
      University of G\"ottingen,
      Germany

}   %% end \Affiliations

\CorrAddress{Manuel Carlan, 
             Department of Statistics, 
             Faculty of Economics,
             University of G\"ottingen, 
             Humboldtallee 3, 
             D--37073 Göttingen, 
             Germany}
\CorrEmail{mcarlan@uni-goettingen.de}
\CorrPhone{}
\CorrFax{}

\Title{Bayesian Discrete Conditional Transformation Models}
\TitleRunning{Bayesian Discrete Conditional Transformation Models}

\Abstract{
We propose a novel Bayesian model framework for discrete ordinal and count data based on conditional transformations of the responses. The conditional transformation function is estimated from the data in conjunction with an a priori chosen reference distribution. For count responses, the resulting transformation model is novel in the sense that it is a Bayesian fully parametric yet distribution-free approach that can additionally account for excess zeros with additive transformation function specifications. For ordinal categoric responses, our cumulative link transformation model allows the inclusion of linear and nonlinear covariate effects that can additionally be made category-specific, resulting in (non-)proportional odds or hazards models and more, depending on the choice of the reference distribution. Inference is conducted by a generic modular Markov chain Monte Carlo algorithm where multivariate Gaussian priors enforce specific properties such as smoothness on the functional effects. To illustrate the versatility of Bayesian discrete conditional transformation models, applications to counts of patent citations in the presence of excess zeros and on treating forest health categories in a discrete partial proportional odds model are presented.
}


\begin{document}

%%% Title page
%%% -------------------------------------------------------------------------------
%%% Use command\maketitle to produce the title page.
\maketitle

%%% Main text
%%% ------------------------------------------
\section{Introduction}
Discrete data commonly occur in almost every scientific area. In this article, we focus on the two relevant cases of count data and ordinal data as special instances of discrete response structures.
Before the advent of generalized linear models \citep[GLM,][]{nelder1972generalized}, the peculiarities of count data were either ignored or treated simply by log transformations \citep{sokal1981principles}.
Then, the standard modeling approach for count data $Y \in \{0,1,2, \ldots\}$ became Poisson regression, $Y|\matr{x} \sim \mathrm{Po}(\lambda_{\matr{x}})$. Since the Poisson distribution often turned out to be too simplistic for many applications, more advanced regression models were introduced as described e.g. by \cite{cameron1998regression}, \cite{winkelmann2008econometric} and \cite{hilbe2011negative} for negative binomial regression, $Y|\matr{x} \sim \mathrm{NB}(\lambda_{\matr{x}}, \nu)$ accounting for potential overdispersion. Generalized additive models \citep[GAM,][]{hastie1990generalized} unify these model types into one framework and drop the linearity assumption for the regression predictor. They require a fixed response distribution that belongs to the exponential family.\\
Similar to counts, ordered categorical data $Y \in \{1, \ldots, c+1\}$ occur in a manifold of scientific disciplines such as medicine or the social sciences. A researcher in medicine for example, may want to distinguish between different kinds of infection grades, while an ecologist could be interested in measuring forest health in terms of defoliation categories. Exploiting the natural ordering in these kind of data is firmly established in the statistical community by cumulative link models as shown in \cite{mccullagh1980regression}. Prominent versions are the discrete proportional odds model and the discrete proportional hazards model \citep{tutz2011regression}. In its simplest form, the cumulative link model is given by
$\pi_{r} = P(Y = r) = F(\gamma_r - \matr{x}^T\bvec) - F(\gamma_{r-1} - \matr{x}^T\bvec), r = 1, \ldots, c+1$
with some pre-specified cumulative distribution function $F$ or equivalently $P(Y \leq r) = F(\gamma_r - \matr{x}^T \bvec) = \pi_{1} + \pi_{2} + \ldots + \pi_{r},$
where $\sum_{r=1}^{c+1} \pi_{r} =1$ is required and the ordering $-\infty \equiv \gamma_0 < \ldots, < \gamma_{c+1} \equiv \infty$ needs to be obliged. It is possible to include category-specific regression effects $\matr{x}^T \bvec_r$ resulting in a (linear) non-proportional odds \citep{peterson1990partial}  or the non-proportional hazards model, depending on the choice of $F$.\\
The dissemination of MCMC simulation techniques led to the development of Bayesian analogues for established models in the form of Bayesian GLMs \citep{dey2000generalized} with many extensions e.g. by \cite{fruhwirth2006auxiliary}, \cite{fruhwirth2009improved} , \cite{rodrigues2003bayesian} and the Bayesian generalized additive model \citep[GAM,][]{brezger2006generalized}. \cite{ghosh2006bayesian} describe a Bayesian treatment of zero-inflated regression models and \cite{klein2015bayesian} introduce zero-inflated and overdispersed count data to the framework of Bayesian structured additive distributional regression \citep{klein2015bayesian2}. In a non-transformation environment, \cite{lavine1995nonparametric} and \cite{dunson2005bayesian}, among others, apply a (strictly) isotonic regression function for count responses on basis of a Dirichlet process mixture prior.\\
To cross the bridge between discrete ordinal and count regression models, we consider count data as ordinal categorical data with a very high number of intercept thresholds which, however, are not estimated but rather fixed by design at all non-negative integers. Methodologically, both approaches are unified by the idea of a direct parametrization of the transformation function. Similar to \cite{ctm} we treat the smooth parametrization of the thresholds as the defining element of the count transformation approach used in this article. While overdispersion is absorbed by the smooth transformation of the counts, we supplement the model with a second component that explicitly accounts for eventual zero inflation. For a discussion of the connection between (binary) regression and transformation models, see \cite{doksum1990correspondence}.

To summarize, in this article, we
\begin{itemize}
\item propose a Bayesian approach for count transformation models based on flexible transformation functions that are inferred from the data which -- in its simplest form with linear covariate shift effects -- results in a distribution-free yet interpretable model framework for count data that automatically accounts for over- and underdispersion in the response distribution,
\item account for excess zeros in two-component mixtures models,
\item propose a Bayesian approach for cumulative link transformation models with Bayesian proportional odds and proportional hazards models as special cases,
\item allow for the inclusion of category-specific effects resulting in non-proportional transformation model types,
\item combine both model types into the class of Bayesian discrete conditional transformation models (BDCTM) and establish it as an extension of Bayesian conditional transformation models (BCTM) for continuous responses,
\item supplement all models with nonlinear, possibly high-dimensional covariate effects and interactions, and
\item illustrate BDCTM's capability in the presence of count and categoric data in two applications.
\end{itemize}

The rest of this article is structured as follows:  Section \ref{sec:bctm} introduces the model class we refer to as Bayesian discrete conditional transformation models with a preliminary discussion of its building blocks. Section \ref{sec:inf} contains a description of posterior estimation. A simulation study evaluating BDCTM's performance in a count data setting is presented in Section \ref{sec:sims}. Section \ref{sec:apps} features an application on patent citation counts and an application on forest health categories. We conclude in Section \ref{sec:end}.

\section{Bayesian Discrete Conditional Transformation Models}\label{sec:bctm}
In what follows, we introduce Bayesian discrete conditional transformation models as a model class that represents a novel approach to the direct estimation of the conditional distribution function $F_{Y|\matr{X}=\matr{x}}(y)$ based on an independent sample of discrete responses $Y_1, \ldots, Y_n$ conditional on covariates $\matr{x}_1, \ldots, \matr{x}_n$. We broadly distinguish between cases of count data and ordered categorical data with a finite sample space, which have to be addressed by different assumptions on the sampling distribution and different basis functions.

Let $y$ be an observation of a count or ordered categorical response variable $Y$ and let $\matr{x}^T= (x_1 , \ldots, x_q)$ be a vector of observed explanatory variables. Moreover, let $F_Z$ be the cumulative distribution function of an a priori chosen reference distribution, linking a discrete and monotonically increasing transformation function $h(\cdot|\matr{x})$ to the conditional distribution function $F_{Y|\matr{X}=\matr{x}}(y|\matr{x})$ via the connection
\begin{align}\label{eq:1}
F_{Y|\matr{X}=\matr{x}}(y|\matr{x}) &= P(Y \leq y|\matr{x}) = F_Z(h( y | \matr{x})).
\end{align}
The responses are transformed towards the reference distribution conditionally on $\matr{x}$ by means of the transformation function $h(y|\matr{x})$.
Through allowing different complexities of the transformation function $h(\cdot|\matr{x})$, BDCTM is able to resemble and expand on established models for count and ordinal data without requiring a fixed response distribution. The encompassing goal of all models described in this paper is to obtain an estimate of the distribution function $F_{Y|\matr{X}=\matr{x}}$ by means of estimating $h( y | \matr{x})$. In contrast to Bayesian CTMs for continuous responses, the transformation function will no longer be bijective since a continuous reference distribution is linked to the CDF of a discrete response variable.

We proceed with discussing each of the components of a BDCTM in more detail. Section~\ref{sec:trafo} introduces the basic structure assumed for the transformation functions. Sections~\ref{sec:counts} and \ref{sec:ordinal} presents model variants for count data and ordinal responses, respectively, while Section~\ref{sec:basis} discusses a generic basis function representation for the transformation functions. Section~\ref{sec:priors} introduces the corresponding prior assumptions, Section~\ref{sec:par_trafo} discusses partial contributions to the transformation function, and Section~\ref{sec:ref} contemplates on the relevance of the choice of the reference distribution.

\subsection{Transformation Functions}\label{sec:trafo}
Similar to \cite{hothorn2014conditional}, we assume an additive decomposition on the scale of the transformation function into $J$ partial transformation functions
\begin{align}\label{eq:add_decomp}
h(y|\bm{x}) &= \sum_{j=1}^J h_j(y|\bm{x}),
\end{align}
where $h_j(y|\bm{x})$ are response-covariate interactions that are monotone only in direction of $y$. We denote partial transformation functions that depend only on the covariates simply by $h(\matr{x})$. A simple transformation model, for example, is obtained by setting $h_1(y|\matr{x})=h_Y(y)$ and $h_2(y|\matr{x})=h(\matr{x})$. We explicitly allow the inclusion of linear and nonlinear covariate effects, i.e.
\begin{align}\label{hx}
h(\matr{x}) &= \sum_{j=1}^J h_j(\matr{x})= \matr{z}^T \bvec + f_{1}(\matr{\nu}) + \ldots + f_{L}(\matr{\nu}),
\end{align}
where in $\matr{x}=(\matr{z}^T,\matr{\nu}^T)^T$, $\matr{z}$ contains all covariates associated with linear effects and $\matr{\nu}$ contains covariates with assumed nonlinear effects $f_1, \ldots, f_L$.

\subsection{Count Transformation Models}\label{sec:counts}
We distinguish between two related model types for count data: simple shift count transformation models that are able to deal with overdispersion and two-component mixture transformation models that can additionally deal with excess zeros.
\paragraph{Mean-shift count transformation models} Regular count transformation models are defined by shifts of the nonlinear baseline transformation function $h_Y$:
\begin{align}\label{eq:shift}
\begin{split}
F_{Y|\matr{X}=\matr{x}}(y |\matr{x}) &= F_Z(h_Y(\lfloor y \rfloor) - h(\matr{x}))
\end{split}
\end{align}
where $\lfloor y \rfloor$ denotes the floor function returning the greatest integer less than or equal to $y$.
Since all moments besides the conditional mean (which is shifted by $h(\matr{x})$) are captured solely by $h_Y(\lfloor y \rfloor)$, independently of the covariates, the resulting model is not affected by over- or underdispersion. Model \eqref{eq:shift} is similar to a regular linear transformation model, but the application of the floor function leads to jumps at the respective integers, such that the transformation function $h_Y(y)$ is only evaluated at the distinctive response values $y \in \{0,1,2,\ldots\}$ and, as a consequence, the overall transformation is no longer invertible. The likelihood-based version of this model type restricted to linear covariate shifts was discussed in detail in \cite{ctm}.
\paragraph{Two-component mixture count transformation models} Besides over- and underdispersion, count data often come with an excess number of zeros, which needs to be accomodated in the model. One possibility is to add a second component to the linear transformation function that captures zeros \citep{hothorn2018most}. A transformation function in that vein can be depicted as:
\begin{align}\label{eq:hurdle}
F_{Y|\matr{X}=\matr{x}}(y |\matr{x}) &= F_Z(h_Y(\lfloor y \rfloor ) - h(\matr{x}) + \mathbbm{1}(y=0)(\beta_0 - h_0(\tilde{\matr{x}}))),
\end{align}
where $h_0(\tilde{\matr{x}})$ and $h(\matr{x})$ can consist of different linear and nonlinear effects of different sets of covariates. This two component mixture transformation model  resembles a hurdle model with hurdle at zero, where the probability of an excess zero is perceived as the mean-shifted deviation from a regular count transformation model at $y=0$:
%\begin{align*}
%P(Y=0|\matr{X}=\matr{x}) &= F(h(0) + \alpha_0 - m(\matr{x}^T(\bvec + \bvec_0)).
%\end{align*}
\begin{align}\label{eq:zero}
P(Y=0|\matr{X}=\matr{x}) &= F_Z(h_Y(0) - h(\matr{x}) +(\beta_0 - h_0(\tilde{\matr{x}}))).
\end{align}
The process generating non-zeros in this case is not explicitly truncated but stems from a transformation function that excludes the zeros.

All count transformation functions of this type have in common that they act on the floor function $\lfloor y \rfloor$, resulting in step functions in direction of $y$ and thus the desired discrete distribution functions. Comparing this to the ordinal response models discussed in the next section, count data transformation models can also be considered as introducing a latent, continuous scale, implicitly determined by the transformation function, with a large number of pre-specified thresholds corresponding to the non-negative integers.

\subsection{Cumulative link transformation models}\label{sec:ordinal}
For ordered categorical data, we distinguish between cumulative models with and without category-specific shifts. From a transformation perspective, the latter are modeled in terms  of response-covariate interactions that can be linear or nonlinear in direction of the respective covariate.
\paragraph{Proportional models} The simplest cumulative transformation model is
\begin{align}\label{eq:prop}
\begin{split}
F_{Y|\matr{X}=\matr{x}}(y_r|\matr{x}) &= F_Z(h_Y(y_r) - h(\matr{x})),
\end{split}
\end{align}
where the term $h(\matr{x})$, which is independent of the category $r$, constitutes the log-odds ratio to $h(0)$ or the log-hazard ratio  in model types \eqref{eq:shift} and \eqref{eq:prop}, depending on the choice of reference distribution.
\paragraph{Non-proportional models} Models of type \eqref{eq:prop} can be generalized by a category-specific shift resulting in the model
\begin{align}\label{eq:non-prop}
\begin{split}
F_{Y|\matr{X}=\matr{x}}(y_r|\matr{x}) &= F_Z(h_Y(y_r) + h_r(\matr{x}))
\end{split},
\end{align}
where $h_r(\matr{x})$ induces the category-specific shifts, resulting in linear or nonlinear non-proportional odds or hazards models depending on $F_Z$ and on whether $h_r(\matr{x})$ consists of linear or nonlinear effects. Partial proportional models as shown in the application in Section \ref{sec:app2} consist of a mixture of proportional and non-proportional effects. The reparameterization illustrated in the following section guarantees that the implied probabilities $P(Y =r) = F_Z(\gamma_r - h_r(\matr{x})) - F_Z(\gamma_{r-1} - h_{r-1}(\matr{x}))$ are always positive.  

\subsection{A generic joint basis}\label{sec:basis}
We assume that each of the $J$ partial transformation functions can be approximated by a linear combination of basis functions $\matr{c}_j$ such that
\begin{align*}
h_j(y|\matr{x})  = \matr{c}_j(y,\matr{x})^T \gvec_j,
\end{align*}
where $\gvec_j$ is a vector of basis coefficients. Based on the additivity assumption in \eqref{eq:add_decomp},
the complete conditional transformation function can be denoted as
\begin{align}\label{eq:full_h}
\begin{split}
h(y|\matr{x}) = \matr{c}(y,\matr{x})^T\gvec
\end{split}
\end{align}
with joint basis 
%\begin{align*}
%\bm{c}(y,\matr{x}) = (\bm{a}_1(y)^T \otimes \bm{b}_1(\matr{x})^T, \ldots, \bm{a}_J(y)^T\otimes \bm{b}_J(\matr{x})^T)^T
%\end{align*}
\begin{align*}
\bm{c}(y,\matr{x}) &= (\bm{c}_1(y, \matr{x})^T, \ldots, \bm{c}_J(y,\matr{x})^T)^T
\end{align*}
and $\gvec$ contains all partial basis coefficient vectors,
\begin{align}
\gvec = (\gvec_1^T, \ldots, \gvec_J^T)^T.
\end{align}
This allows us to write all discrete conditional transformation models treated in this article in the general form
\begin{align}\label{full_bdctm}
F_{Y|\matr{X}=\matr{x}}(y) &= F_Z(\matr{c}(y,\matr{x})^T \gvec).
\end{align}
We call models of type \eqref{full_bdctm} Bayesian discrete transformation models (BDCTM). They can be conceived as extensions of the versatile model class of Bayesian conditional transformation models (BCTM) for continuous responses that were introduced by \cite{bctm}, taking the additional challenges arising from discrete responses into account. In this tradition, a BDCTM is fully specified by a reference distribution $F_Z$, the joint basis $\matr{c}(y,\matr{x})$ and a vector of basis coefficients $\gvec$ together with suitable priors which are introduced in the next section. The rest of this section discusses the generic basis that is used by the BDCTM in greater detail.

Let $\matr{a}_j$ denote a basis transformation of $y$ with dimension $D_1$, collecting evaluated basis functions $B_{j1d_1}(y),d_1=1,\ldots, D_1$ and let $\matr{b}_j$ denote a basis transformation of $x$ with dimension $D_2$ collecting evaluated basis functions $B_{j2d_2}(x),d_2=1,\ldots, D_2$. The resulting effects are approximated by the linear combinations
\begin{align*}
h_j(y)=\sum_{d_1=1}^{D_1} \gamma_{j1d_1}B_{j1d_1}(y)=\matr{a}(y)^T \gvec_{j1}, \quad h_j(x)=\sum_{d_2=1}^{D_2} \gamma_{j2d_2}B_{j2d_2}(x) = \matr{b}_j(x)^T\gvec_{j2},
\end{align*}
where $\gvec_{j1}= (\gamma_{j11}, \ldots, \gamma_{j1D_1})^T$ and  $\gvec_{j2}= (\gamma_{j21}, \ldots, \gamma_{j2D_2})^T$ are partially reparameterized versions of the vectors of corresponding basis coefficients $\bvec_{j1}$ and $\bvec_{j2}$.
The conditional transformation approach commonly involves response-covariate interactions (see e.g. model types \eqref{eq:zero} and \eqref{eq:non-prop}) which is why we parametrize each partial transformation function generically as
\begin{align}\label{eq:full_b}
\begin{split}
h_j(y |x) &= \matr{c}_j(y, x)^T \gvec_j = (\matr{a}_j(y)^T \otimes \matr{b}_j(x)^T)^T \gvec_j \\
&= \sum_{d_1=1}^{D_1} \sum_{d_2=1}^{D_2} \gamma_{j,d_1d_2} B_{d_1}(y)B_{d_2}(x),
\end{split}
\end{align}
where the Kronecker product forms parametric interactions between the evaluated basis functions and $\gvec_j$ is a basis vector of dimension $D=D_1 D_2$.
A collection of special cases can be found in Section \ref{sec:par_trafo}.

We require all transformation functions to be strictly monotonically increasing solely in the direction of $y$ but not in direction of the explanatory variables such that $F_{Y|\matr{X}=\matr{x}}(y_{j}|\matr{x}) < F_{Y|\matr{X}=\matr{x}}(y_{j+1}|\matr{x})$ for all $y_j < y_{j+1}$. This property needs to be accomodated in the basis. For this, we adopt the approach of \cite{pyawood} for monotonically increasing smooth functions. The vector $\gvec_j$ is reparameterized as $\gvec_j= \matr{\Sigma}_j \bt_j$, where $\matr{\Sigma}_j = \matr{\Sigma}_{D_1} \otimes \matr{I}_{D_2}$ and $\matr{\Sigma}_{D_1}$ is given by the lower triangular matrix of size $D_1$ such that $\Sigma_{D_1,kl}=0$ if $k < l$ and $\Sigma_{D_1,kl} = 1$ if $k \geq l$. The vector $\bt_j$ of dimension $D=D_1 D_2$ contains a mixture of unexponentiated and exponentiated $\beta$-coefficients given by
\begin{align}\label{eq:bt2}
\bt_j = (\beta_{j,11}, \ldots, \beta_{j,1D_2}, \exp(\beta_{j,21}), \ldots \exp(\beta_{j,2D_2}), \ldots , \exp(\beta_{j,D_1D_2}))^T.
\end{align}
and $\matr{I}_{D_2}$ is an identity matrix of size $D_2$. An unconditional transformation function $h_Y(y)$ is obtained by setting $D_2=1$ and a function of type $h(\matr{x})$ is obtained by setting $D_1=1$.

The vector of basis coefficients for the whole conditional transformation function $h(y|\matr{x})$ is given by $\gvec =\matr{\Sigma}\bt$ where $\bt =(\bt_1^T, \ldots, \bt_J^T)^T$ is based on $\bvec = (\bvec_1^T, \ldots, \bvec_J^T)^T$. Matrix $\matr{\matr{\Sigma}}$ is block diagonal with $\matr{\Sigma}_j$ as diagonal elements.

Of course, other basis specification could be employed to set up BDCTMs, as long as monotonicity along $y$ is ensured. For example, the increasing splines considered in continuous ordinal regression \citep{OrdinalRegressionModelsforContinuousScales} would be a potential alternative. We rely on Bayesian P-splines and their tensor product interactions since these have been extensively studied in Bayesian structured additive regression and enable efficient and stable computations.

\subsection{Priors}\label{sec:priors}
We adopt the principle of Bayesian P-splines \citep{lang2004bayesian} and assume partially improper multivariate Gaussian priors for the unconstrained vectors $\bvec_{j1}$ and $\bvec_{j2}$ (the reparameterized vectors $\gvec_{j1}$ and $\gvec_{j2}$ are based on) such that
\begin{align}\label{eq:mvnprior}
\begin{aligned}
p(\bvec_{j1} | \tau_{j1}^2) &\propto \left(\frac{1}{\tau_{j1}^2}\right)^\frac{\mathrm{rk}(\kmat_{j1})}{2 \tau_{j1}^2} \exp \left(- \frac{1}{2 \tau_{j1}^2} \bvec_{j1}^T \kmat_{j1}\bvec_{j1}\right),\\
p(\bvec_{j2} | \tau_{j2}^2) &\propto \left(\frac{1}{\tau_{j2}^2}\right)^\frac{\mathrm{rk}(\kmat_{j2})}{2 \tau_{j2}^2} \exp \left(- \frac{1}{2 \tau_{j2}^2} \bvec_{j2}^T \kmat_{j2}\bvec_{j2}\right),
\end{aligned}
\end{align}
 where  $\tau_{j1}^2$ and $\tau_{j2}^2$ are marginal smoothing variances, $\text{rk}(\cdot)$ is the rank of a matrix and $\bm{K}_{j1}$ and $\bm{K}_{j2}$ are potentially rank deficient prior precision matrices. The generic formulation of the precision matrix associated with $\gvec_j$ is given by
\begin{align*}
\kmat_{j}=\frac{1}{\tau_{j1}^2}(\matr{K}_{j1} \otimes \matr{I}_{D_2}) + \frac{1}{\tau_{j2}^2}(\matr{I}_{D_1} \otimes \matr{K}_{j2}),
\end{align*}
where precision matrices $\matr{K}_{j1}$ and $\matr{K}_{j2}$ control the penalty in the direction of $y$ and $x$ respectively. For unconditional transformation functions or pure covariate functions, $\matr{K}_{j1}$ and $\matr{K}_{j2}$ are respectively set to $\matr{0}$ such that only the prior precision of the corresponding effect is used. Specific choices are discussed in the next section. The model precision matrix $\kmat$ is given as the block diagonal matrix with matrices $\kmat_{j}$ as diagonal elements.

The smoothing variances $\tau_{j1}^2$ and  $\tau_{j2}^2$ are associated with inverse gamma priors, $\tau_{j1}^2 \sim \mathrm{IG}(a_{j1},b_{j1})$ and $\tau_{j2}^2 \sim \mathrm{IG}(a_{j2},b_{j2})$. All model parameters are collected in $\bm{\vartheta} = ( \bvec_1, \ldots, \bvec_J, \tau_{11}^2,\tau_{12}^2 \ldots, \tau_{J1}^2, \tau_{J2}^2)^T$ with joint prior $p(\bm{\vartheta})$ 

\subsection{Partial transformations}\label{sec:par_trafo}
We start this section by introducing the two types of basis functions we use in $\matr{a}$, depending on whether $Y$ is a count variable or discrete ordinal followed by a brief discussion of choices for $\matr{b}$ together with suitable precision matrices.
\paragraph{Smooth basis for count transformations} In case of a count response $Y \in \{0,\ldots\}$, $\matr{a}$ consists of B-spline basis functions $B_{d_1}$ i.e. $\matr{a}_j(y) =(B_1(y), \ldots, B_{D_1}(y))^T$. It may be useful to parametrize the transformation function on the log-scale, i.e $\matr{a}_j(\log(y))$ or $\matr{a}_j(\log(y+1))$, where especially the latter can be beneficial numerically if there are many small and some large counts.
Smooth monotonic effects of a count transformation subject to the reparameterization in \eqref{eq:bt2} are supplemented with a penalty matrix $\kmat_{j1}=\matr{D}_1^T\matr{D}_1$ based on a $(D_1-2) \times D_1$ partial first-difference matrix $ \matr{D}_1$ that is zero except that $D_{i,i+1}=-D_{i,i+2}=1$ for $i=1, \ldots, D_1-2$  to achieve shrinkage towards a straight line \citep{pyawood}.
\paragraph{Discrete basis for ordinal categorical data} For ordered categorical responses, $Y \in \{1, \ldots, c+1\}$ we assign one parameter to each category except for the reference category $c+1$ \citep{hothorn2018most}. As a basis, we use the unit vector $\matr{e}_c$ of length $c$, i.e. $\matr{a}_j(y_r) = \matr{e}_c(r)$, where
\begin{align}\label{eq:3}
\begin{aligned}
Y &= r \Longleftrightarrow \boldsymbol{e}_c(r) = (0, \ldots, 1, \ldots, 0)^T ,&r=1,\ldots, c.
\end{aligned}
\end{align}
The corresponding precision matrix is $\kmat_{j1} = \matr{0}$.

\paragraph{Bases for covariates effects} For covariate effects we allow linear bases $\matr{b}_j(\matr{z})= (z_1, \ldots, z_p)^T$ together with precision matrix $\kmat_{j2}=\matr{0}$ and B-spline bases for nonlinear effects $\matr{b}_j(\nu)=(B_1(\nu), \ldots, B_{D_2}(\nu))^T$ with a second order random-walk precision matrix $\kmat_{j2}$. All bases involving B-spline basis functions can be centered around zero for identification purposes.

Transformation random effects $h_j(x) = \beta_g$ are based on the grouping indicator $g \in \{1, \ldots, G\}$. The corresponding $G$-dimensional basis vector $\boldsymbol{b}_{j}(g)$ has entry one if $x$ belongs to group $g$ and zero otherwise. We set $\boldsymbol{K}_j = \boldsymbol{I}_G$ for i.i.d.\ random effects.
Regular non-monotonic tensor splines as used in the forest health application in Section~\ref{sec:app2} can be retrieved by using the specification in (\ref{eq:full_b}) and setting $\boldsymbol{\gamma}_j = \boldsymbol{\beta}_j$.

\subsection{Reference distribution}\label{sec:ref}
In the context of discrete conditional transformation models, the reference distribution function $F_Z$ plays the role of the inverse link function controlling the interpretational scale of the impact of the explanatory variables. While it can be chosen arbitrarily in theory, we concentrate on distributions with log-concave densities for $F_Z$ to guarantee uniqueness of the maximum likelihood estimate, which usually will also imply unimodality of the posterior. Furthermore, it is advised to consider characteristics such as right-skewness or the support of the count data distribution in the selection process. Prominent choices for $F_Z$ are
\begin{itemize}
\item $F_\text{SL}(z)=(1 + \exp(-z))^{-1}$, i.e. the standard logistic distribution,
\item $\Phi(z)$, i.e. the standard normal distribution, and
\item $F_\text{MEV}(z)= 1- \exp(-\exp(z))$, i.e. the minimum extreme value distribution
\end{itemize}
resulting in logit, probit or cloglog interpretations of the covariate effects. Setting $F_Z=F_{\text{SL}}$ for example results in the discrete proportional odds model and $F_Z=F_{\text{MEV}}$ results in the proportional hazards model with $h(\matr{x})$ becoming the log-odds ratio or the log-hazard ratio to $h(\matr{0})$, respectively \citep{hothorn2018most}.

To reflect specific properties of the data generating process, other link functions that have been considered in the context of generalized linear models, such as skew-logistic or t-distributed link functions to reflect strong asymmetry or heavy tails may be considered. However, given the flexibility of the transformation function, we do not expect large gains from such specifications since both asymmetry and tail behaviour should be taken up by the transformation function, leaving only small potential for improving the fit via the link function. We therefore suggest to stick to the defaults and to select the reference distribution according to preferences on model interpretation.

\subsection{Transformation probability mass functions}
In this section, we introduce the transformation probability mass functions (PMFs) resulting from the different sampling assumptions that come with count and ordinal categoric data as well as the resulting transformation likelihoods. To emphasize that $\gvec$ are partially nonlinear reparameterizations of $\bvec$, we write $\gvec(\bvec)$.
Following \cite{hothorn2018most}, the log-transformation PMF of a conditionally independent (count) response $Y$ with unbounded support $Y \in \{0,1,\ldots, \}$ is given by
%\begin{align*}
%f(y_k |h) &=\begin{cases}
%    = F(\matr{a}(0)^T\bt - \matr{x}_i^T\bvec) & y_i=0\\
%    = F(\matr{a}(y_i)^T\gvec - \matr{x}_i^T\bvec) - F(\matr{a}(y_i - 1)^T\gvec - \matr{x}_i^T\bvec) & y_i >1.
%    \end{cases}.
%\end{align*}
\begin{align}\label{disc_ll1}
\log(f_Z(y |\bvec)) &=\begin{cases}
     \log [F_Z(\matr{c}(y_k,\matr{x})^T \gvec(\bvec))] & k=1\\
     \log[F_Z(\matr{c}(y_k,\matr{x})^T \gvec(\bvec)) - F_Z(\matr{c}(y_{k-1},\matr{x})^T \gvec(\bvec)] & k > 1.
    \end{cases}
\end{align}
In case of an ordinal categorical response with bounded support $Y\in\{y_1,\ldots,y_{c+1}\}$, the corresponding conditional distribution function needs to take the additional constraint for the reference category $c+1$, $P(Y \leq y_{c+1} |\matr{X}=\matr{x})=F_Z(h(y_{c+1}|\matr{X}=\matr{x}))=1$ into account. The transformation PMF is then given by
\begin{align}\label{disc_ll2}
f_Z(y |\bvec)= 
\begin{cases}
     [F_Z(\matr{c}(y_k,\matr{x})^T \gvec(\bvec)))] & k =1\\
     [F_Z(\matr{c}(y_k,\matr{x})^T \gvec(\bvec)) - F_Z(\matr{c}(y_{k-1},\matr{x})^T \gvec(\bvec))] & k = 2, \ldots, c\\
     [1 - F_Z(\matr{c}(y_c,\matr{x})^T \gvec(\bvec))] & k = c+1.\\
    \end{cases}
\end{align}
With the convention $F_Z(h(y_0))=F_Z(h(-\infty))=0$ and $F_Z(h(y_{c+1}))=F_Z(h(\infty))=1$, the conditional PMF simplifies to 
\begin{align}\label{eq:trafo_dens}
f_Z(y_k|\bvec) = F_Z(\matr{c}(y_k,\matr{x})^T \gvec(\bvec) -F_Z(\matr{c}(y_{k-1},\matr{x})^T \gvec(\bvec))
\end{align}
encompassing count and ordered categoric models in a unified framework \citep{hothorn2018most}. Based on \eqref{eq:trafo_dens}, the transformation log-likelihood for independent observations $(y_i, \matr{x}_i), i =1, \ldots, n$ is given by
\begin{align*}
l(\bvec) =\sum_{i=1}^n \log(F_Z(\matr{c}(y_i,\matr{x}_i)^T \gvec(\bvec)) - F_Z(\matr{c}(y_i -1,\matr{x}_i)^T \gvec(\bvec))). 
\end{align*}
The likelihood is chosen according to the discrete response structure only, while the transformation function determines whether excess zeros are accounted for or if the category-specific effects are included for example.
With all building blocks in mind, a BDCTM can be fully specified by the set $\lbrace  \matr{\vartheta}|F_Z,\matr{c}, \pi_\vartheta(\cdot)\rbrace$ of unknown model parameters  $\matr{\vartheta}$ given a choice for the basis $\matr{c}$, the reference distribution $F_Z$ and joint prior $\pi_\vartheta$ \citep{bctm}.

\section{Posterior Inference}\label{sec:inf}
For Bayesian inference, we rely on Markov chain Monte Carlo (MCMC) simulation techniques. We sketch the most relevant parts of the algorithm in this section.
\paragraph{Update of the basis coefficients}  The log-full conditional of the basis coefficients with injected proportionality is given by
\begin{align*}
\log(p(\bvec | \cdot)) \propto l(\bvec) - \frac{1}{2}\bvec^T \kmat \bvec,
\end{align*}
where the second term arises from the multivariate Gaussian prior. The gradient of the unnormalized log-posterior is needed for inference and is given by
\begin{align*}
\matr{s}(\bvec) &= \sum_{i=1}^n\frac{f_Z(\matr{c}(y_i,\matr{x}_i)^T \matr{\Sigma} \bt)\matr{c}(y_i,\matr{x}_i)^T\matr{\Sigma}\matr{C} - f_Z(\matr{c}(y_i-1,\matr{x}_i)^T \matr{\Sigma} \bt)\matr{c}(y_i-1,\matr{x}_i)^T\matr{\Sigma}\matr{C}}{F_Z(\matr{c}(y_i ,\matr{x}_i)^T \matr{\Sigma} \bt) - F_Z(\matr{c}(y_i -1,\matr{x}_i)^T \matr{\Sigma} \bt)} - \matr{K}\bvec,
\end{align*}
where $\matr{C}$ is a diagonal matrix of size $D$ with entries
\begin{align*}
C_{dd} &=\begin{cases}
     1 & \text{if } \tilde{\beta}_d = \beta_d\\
    \exp(\beta_d), &\text{otherwise}.
    \end{cases}
\end{align*}
Strong dependencies among the variables (which is partly due to the monotonicity restriction) complicate sampling from the posterior distribution. This is further impeded by the mixed linear-non-linear dependence of the transformation function on $\bvec$ and $\bt$, respectively. Therefore, we use the No-U-turn Sampler \citep[NUTS,][]{hoffman2014no} with dual averaging \citep{Nesterov2009primal} for efficient exploration of the target distribution. The adaptive and dynamic nature of NUTS enables a streamlined estimation process that abolishes the need for costly preliminary tuning runs (needed for setting the number of leapfrog steps and the step size parameter) at the expense of some computation time per iteration. In the following, we distinguish between the burn-in period which determines the number of samples that gets thrown out at the beginning of a Markov chain and the warm-up period which controls the length of the adaptive phase of the algorithm.
Due to the high dependence between parameter blocks, all basis coefficients are updated in one step followed by successive updates of the smoothing variances.
\paragraph{Update of the smoothing variances} In the univariate case, updating the smoothing variance is straightforward by using the full-conditional
\begin{align*}
\tau_{j}^2 | \cdot & \sim \mathrm{IG}\left(a_{j} + \frac{\mathrm{rk}(\bm{K}_{j})}{2}, b_{j} + \frac{1}{2} \bvec_j^T \bm{K}_{j} \bvec_j \right),
%\tau_{j2}^2 | \cdot & \sim \mathrm{IG}\left(a_{j2} + \frac{\mathrm{rk}(\bm{K}_{j2})}{2}, b_{j2} + \frac{1}{2} \bvec_j^T \bm{K}_{j \tau_2^2} \bvec_j \right),
 \end{align*}
where $\kmat_j$ is specified as shown in Section \ref{sec:par_trafo}. However, in case of tensor splines based on a multivariate Gaussian prior with precision matrix
\begin{align}\label{eq:K_tensor}
    \frac{1}{\tau_{j1}^2} (\matr{K}_{j1} \otimes \matr{I}_{D_2}) +\frac{1}{\tau_{j2}^2}(\matr{I}_{D_1} \otimes \matr{K}_{j2}),
\end{align}
we need to consider the generalized determinant of \eqref{eq:K_tensor} when updating the smoothing variances. This aggravates sampling, which is why we introduce an anisotropy parameter $\omega_j \in (0,1)$, resulting in an alternative representation of the precision given by
\begin{align*}
    \frac{1}{\tau_j^2} \kmat_j = \frac{1}{\tau_j^2} \left[\omega_j (\matr{K}_{j1} \otimes \matr{I}_{D_2}) + (1 -\omega_j) (\matr{I}_{D_1} \otimes \matr{K}_{j2}) \right],
\end{align*}
where $\omega_j$ controls how much prior information is assigned to each of the two covariates of the tensor spline. For the BDCTM, we consider a discrete prior for $\omega_j$ which allows to pre-compute a finite set of generalized determinants that can be used within the MCMC simulations. See \cite{kneib2019modular} for a detailed explanation of this approach.

In the following, the hyperparameters of the inverse gamma prior are set to $a_{j1}=a_{j2}=1, b_{j2}=b_{j2}=0.001$ resulting in good and stable performance in all investigated cases.

\paragraph{Numerical stability} \cite{klein2015bayesian} observed numerical problems, if zero-inflation was wrongfully assumed when in fact e.g. a simple Poisson model was due. One reason is  that the estimated predictor for the probability of an extra zero tends towards minus infinity in log-space. This is usually not an issue in models of type \eqref{eq:hurdle} as the coefficients that are related to the zero component are not $\exp$-transformed. In cumulative models with category-specific effects however, flat sections can lead to divergent transitions in which case weakly identified coefficients have to be dropped from the model \citep{pyawood}. This issue can be remedied by adding $\epsilon = 10e^{-6}$ to the diagonal of the precision matrix in this case. Moreover, the target acceptance rate can be increased to up to $.99$ to keep transitions in check.

\paragraph{Software}
All computations were carried out in R version 4.1.0 \citep{rlang}. To improve computation time, likelihoods and score functions were implemented via the package \texttt{Rcpp} \citep{eddelbuettel2011rcpp}. The mass matrix adaption scheme was adopted from  \texttt{adnuts} \citep{adnuts}.

\section{Simulation study}\label{sec:sims}
In this section, we present a simulation experiment that highlights the possible advantages of the count transformation approach in general and compares our Bayesian approach with the likelihood-based linear count transformation model by  \cite{ctm}.

Count transformation models can mimic most well-known models for count data. Therefore, a meaningful simulation study in this setting needs to consider the sensitivity of the flexible transformation function with respect to the true data generating process. In other words, it needs to investigate to what extent the flexible transformation function is able to accommodate eventual overdispersion and other characteristics of possibly complex data generating processes.
\paragraph{Simulation design}
We use a similar simulation design to \cite{ctm} with the following properties:
\begin{itemize}
\item One covariate is generated via $z \sim \mathbb{U}[0,1]$.
\item Conditional on $z$, we consider five different count DGPs:
\begin{itemize}
\item $\mathit{Poisson}$ with mean and variance $\mathbb{E}(Y|z) = \mathbb{V}(Y|z) = \exp(1.2 + 0.8z)$,
\item $\mathit{Negative Binomial}$ with $\mathbb{E}(Y|z) = \exp(1.2 + 0.8z)$ and variance $\mathbb{V}(Y|z) = \mathbb{E}(Y|z) + \mathbb{E}(Y|z)^2/3$, and
\item three different count data generating processes according to $F_Z(\matr{a}_{(8)}( \log(y+1))^T\gvec  - z\beta)$, $\beta=0.8$ with the reference functions $F_Z=F_{\text{SL}}$ ($\mathit{logit}$), $F_Z=\Phi$ ($\mathit{probit}$),
 $F_Z=F_{\text{MEV}}$ ($\mathit{cloglog}$).
\end{itemize}
\item Each dataset was estimated by their corresponding true (oracle) models, i.e. a Poisson GLM ($\mathit{mp}$), a negative binomial GLM ($\mathit{mnb}$), BDCTMs ($\mathit{bmlo}$ denotes the logistic model, $\mathit{bmpr}$ denotes the probit model and $\mathit{bmcll}$ denotes the cloglog model) and a frequentist count transformation model  \citep{ctm} implemented in the \textbf{R}-package \texttt{cotram} \citep{siegfriedcount}, where $\mathit{mlo}$ stands for the logistic model, $\mathit{mpr}$ stands for the probit model and $\mathit{mcll}$ for the cloglog model. Each model type was estimated each DGP, resulting in  $5 \times 8 =40$ models in total.
\item Training and validation sample sizes are set to $250$ and $750$, respectively.
\item The simulation experiment was repeated in $100$ replications with a total iteration number of $2000$ and a burn-in and warm-up phase of length $1000$ such that $1000$ iterations are being used for computing the estimates.
\end{itemize}
\begin{figure}
\centering  
   %\makebox[0pt]{\includegraphics[scale=0.9]
      \makebox[0pt]{\includegraphics[width=\textwidth]
   {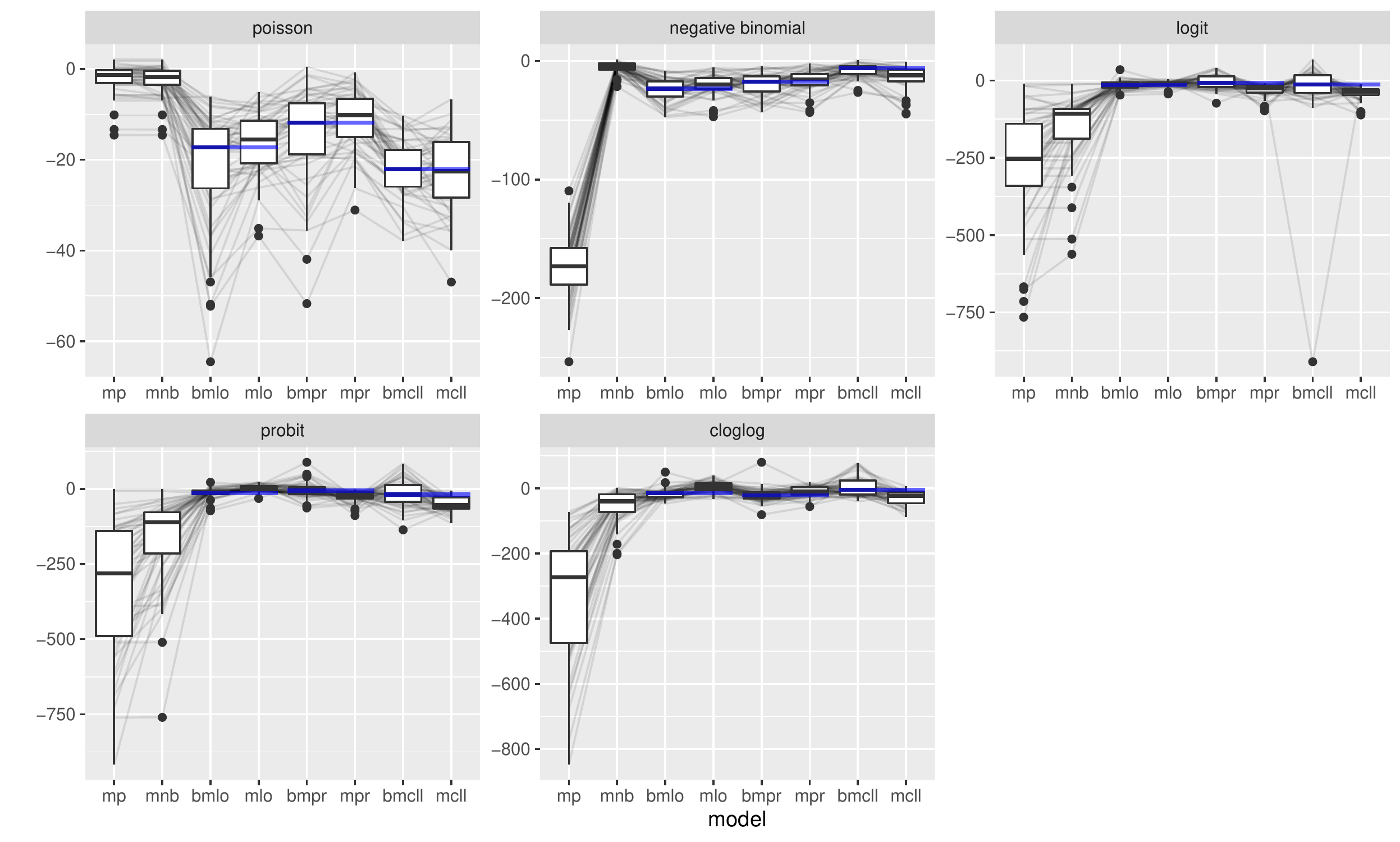}}\caption{Comparison of count data-generating processes on basis of centered out-of-sample log-likelihoods obtained from the respective model. Larger out-of-sample log-likelihoods indicate a better performance of the corresponding model.}\label{fig:sims_all}
   \end{figure}
Each model fit is quantified by means of the centered out-of-sample log-likelihood resulting from the difference between the out-of-sample log-likelihoods of the models and the out-of-sample log-likelihoods of the true data-generating processes evaluated on a hold-out sample, taking a predictive perspective that implicitly controls for differenes in complexity between the models. The results presented in Fig. \ref{fig:sims_all} confirm most of the findings of \cite{ctm} regarding the merits of the count transformation approach.

Based on these results, we can make the following statements:  
\begin{itemize}
\item The Poisson model, being the most rigid model, shows the worst performance with respect to the out-of-sample log-likelihood, if misspecified.
\item As expected, the negative binomial model performs well for the Poisson and the overdispersed case, but shows inferior performance in the remaining scenarios.
\item The fit of both the BDCTM and the \texttt{cotram} model is robust for all considered DGPs, effectively redeeming the promise of providing a flexible model framework for count data that is applicable in many situations.
\item The BDCTM seems to perform better than \texttt{cotram} in the more complicated scenarios and worse especially in settings where a simple Poisson model would be due; this may be less surprising considering BDCTM's spline-based nature in comparison to \texttt{cotram}'s use of Bernstein polynomials.
\end{itemize}
The simulation study confirms the robustness of the BDCTM in the presence of different data generating processes. Its fit is satisfactory in all investigated cases and highly competitive in the more complicated scenarios. While the Poisson distribution only works well in simple scenarios, the negative binomial distribution also works quite well for most scenarios (except the Poisson case). Still, BDCTMs outperform negative binomial regression uniformly over all but the Poisson and the negative binomial scenario.

\section{Applications}\label{sec:apps}
We illustrate possible applications of the BDCTM in this section. For better readability, we add the number of basis functions to the basis, e.g. $\matr{a}_{(q)}$. Code required for reproducing the following applications is openly accessible\footnote{Source code available at \url{https://github.com/manucarl/bdctm_showcase}.}.

\subsection{Patent citations with excess zeros}\label{sec:apps1}
Similar to an author of a scientific publication, an inventor who applies for a patent has to cite all existing patents her work is based on. We analyze the citation number ($\mathit{ncit}: y$) of patents granted by the European Patent Office (EPO).  The considered dataset includes five dummys and three continuous variables. The available continuous covariates are the grant year ($\mathit{year}$), the number of the designated states ($\mathit{ncountry}$) and the number of patent claims ($\mathit{nclaims}$). For a full description of the explanatory variables in the data set of $n=4,805$ observations, see \cite{jerak2006modeling}. A high rate of zeros ($\approx 46\%$) and a big spread  $\text{ncit}=\{0,\ldots, 40\}$ hint on the presence of zero-inflation and overdispersion. A rigorous investigation of this presumption has to consider whether this is holds conditional on the covariates. We let the sampler run for $2000$ iterations with a burn-in and warm-up phase of length $1000$ such that $1000$ iterations are obtained for inference.

\begin{figure}
\centering  
   %\makebox[0pt]{\includegraphics[scale=0.9]
      \makebox[0pt]{\includegraphics[width=\textwidth]
   {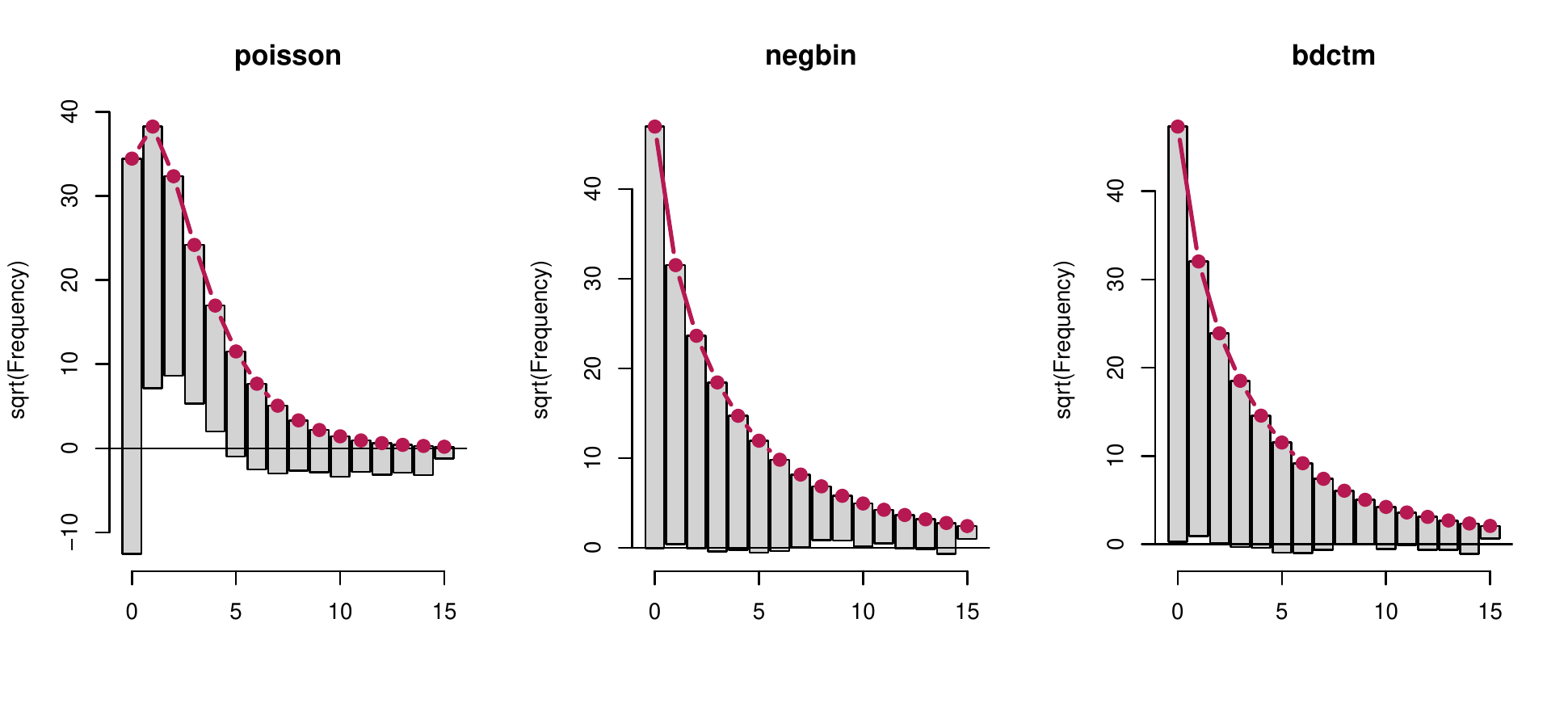}}\caption{\footnotesize Patent citations. Rootograms of the linear Poisson, the linear negative binomial and the simple linear BDCTM model. }\label{fig:patent_rootogramm}
   \end{figure}

We start our investigation with the simple linear transformation model ($\text{BDCTM}_\text{lin}$):
 \begin{align}\label{eq:patent_reg}
F_{\text{SL}}( \matr{a}_8(\log(\lfloor y+1\rfloor))^T\gvec - \matr{z}^T \bvec),
 \end{align}
where the linear predictor $\matr{z}^T \bvec$ contains all available covariates.
As a first in-sample assessment of the practical capabilities of our transformation approach, we want to inspect to what extend the observed frequencies $\text{obs}_r = \sum_{i=1}^n \mathbbm{1}(y_i=r)$ in the data set match the expected frequencies $\text{exp}_r = \sum_{i=1}^n(r;\hat{\gvec}_i)$ derived from the model. Figure \ref{fig:patent_rootogramm} displays the rootograms as introduced by \cite{kleiber2016visualizing} obtained from the model in equation \eqref{eq:patent_reg}, from a Poisson and from a negative binomial GLM with all covariates included in the predictors. Rootograms make use of a horizontal reference line (at zero) to highlight the discrepancies between observed and expected frequencies. The Poisson model clearly underfits the zeros and exhibits an undulating pattern, overpredicting counts between $1$ and $4$ and underpredicting the rest, which is a sign of substantial overdispersion. The flexible transformation function of BDCTM is able to emulate the overdispersion-robust negative binomial model, which is reflected in the bars being closely aligned with the $x$-axis.
In summary, this first visual inspection of the goodness-of-fit confirms that BDCTM is able to ameliorate the impact of overdispersion on the model fit.

\begin{figure}
\centering  
   %\makebox[0pt]{\includegraphics[scale=0.9]
      \makebox[0pt]{\includegraphics[width=\textwidth]
   {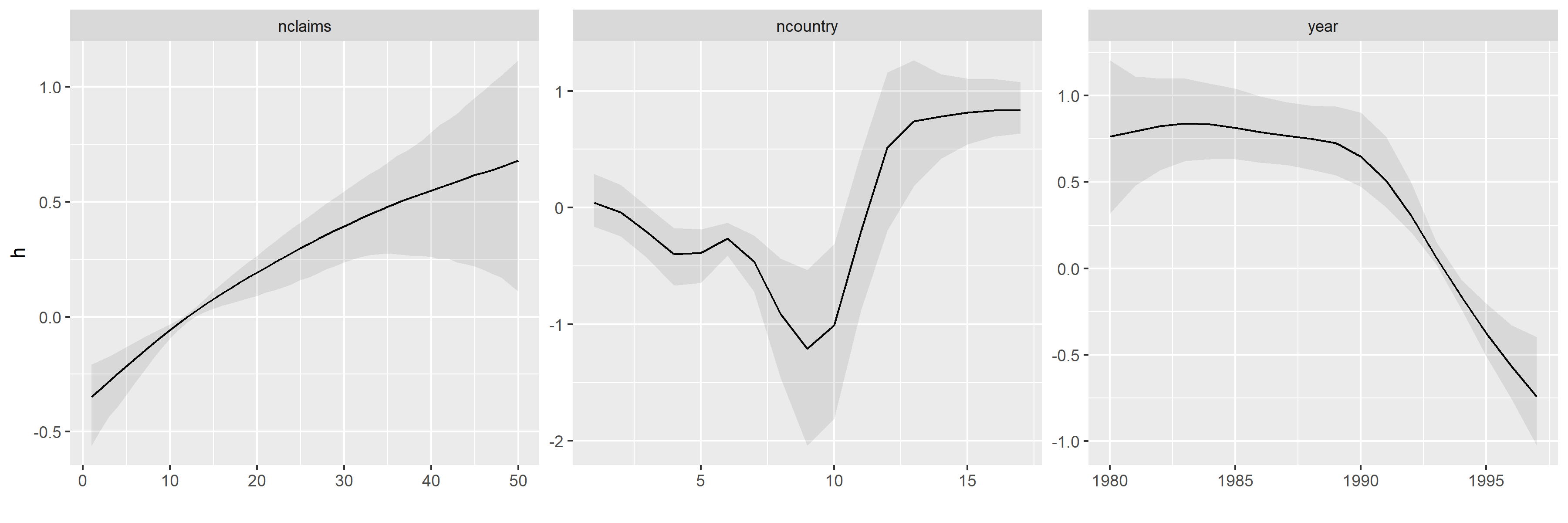}}\caption{\footnotesize Patent citations. Posterior mean estimates of the effects of nclaims, ncountry and year on the log-odds ratio, together with 95\% credible intervals. Remaining covariates are held constant at their mean or are set to zero in case of dummy variables. Estimates belong to $\text{BDCTM}_\text{nl}$.}\label{fig:patent_nl}
   \end{figure}

We also want to pursue the assumption of excess zeros. For this, we consider a two component model ($\text{BDCTM}_\text{hurdle-lin}$) in the vein of \eqref{eq:hurdle} with $h(\matr{x}) = h_0(\matr{x})=\matr{z}^T\bvec$:
\begin{align*}
F_{\text{SL}}( \matr{a}_8(\log(\lfloor y+1\rfloor))^T\gvec - \matr{z}^T \bvec+ \mathbbm{1}(y=0)(\beta_0 - \matr{z}^T\bvec)),
\end{align*}
where again, $\matr{z}$ contains all explanatory variables in the data set. As GLM analogues, we consider the zero-inflated versions of the Poisson and of the negative binomial models. Previous analyses of the data set revealed that assuming nonlinear relationships for the continuous covariates can improve the estimation results \citep{klein2015bayesian}. This does not automatically hold for BDCTM, because the explanatory variables impact the response on a different scale (the scale of the transformation). Therefore, we estimated models of type \eqref{eq:shift} and \eqref{eq:hurdle}, while replacing the covariate functions with additive functions of type \eqref{hx}, i.e. $h(\matr{x})=h_0(\matr{x})= \matr{z}^T\bvec + f(\mathit{ncountry}) + f(\mathit{year}) + f(\mathit{nclaims})$, where $\matr{z}$ now only contains the discrete covariables. In what follows, we refer to these partially nonlinear models as $\text{BDCTM}_\text{nl}$ and $\text{BDCTM}_\text{hurdle-nl}$, respectively. Fig. \ref{fig:patent_nl} shows the estimated nonlinear effects of $\mathit{ncountry}, \mathit{year}$ and $\mathit{nclaims}$ on the log-odds ratio from model $\text{BDCTM}_\text{hurdle-nl}$.

\begin{figure}
\centering  
   %\makebox[0pt]{\includegraphics[scale=0.9]
      \makebox[0pt]{\includegraphics[width=\textwidth]
   {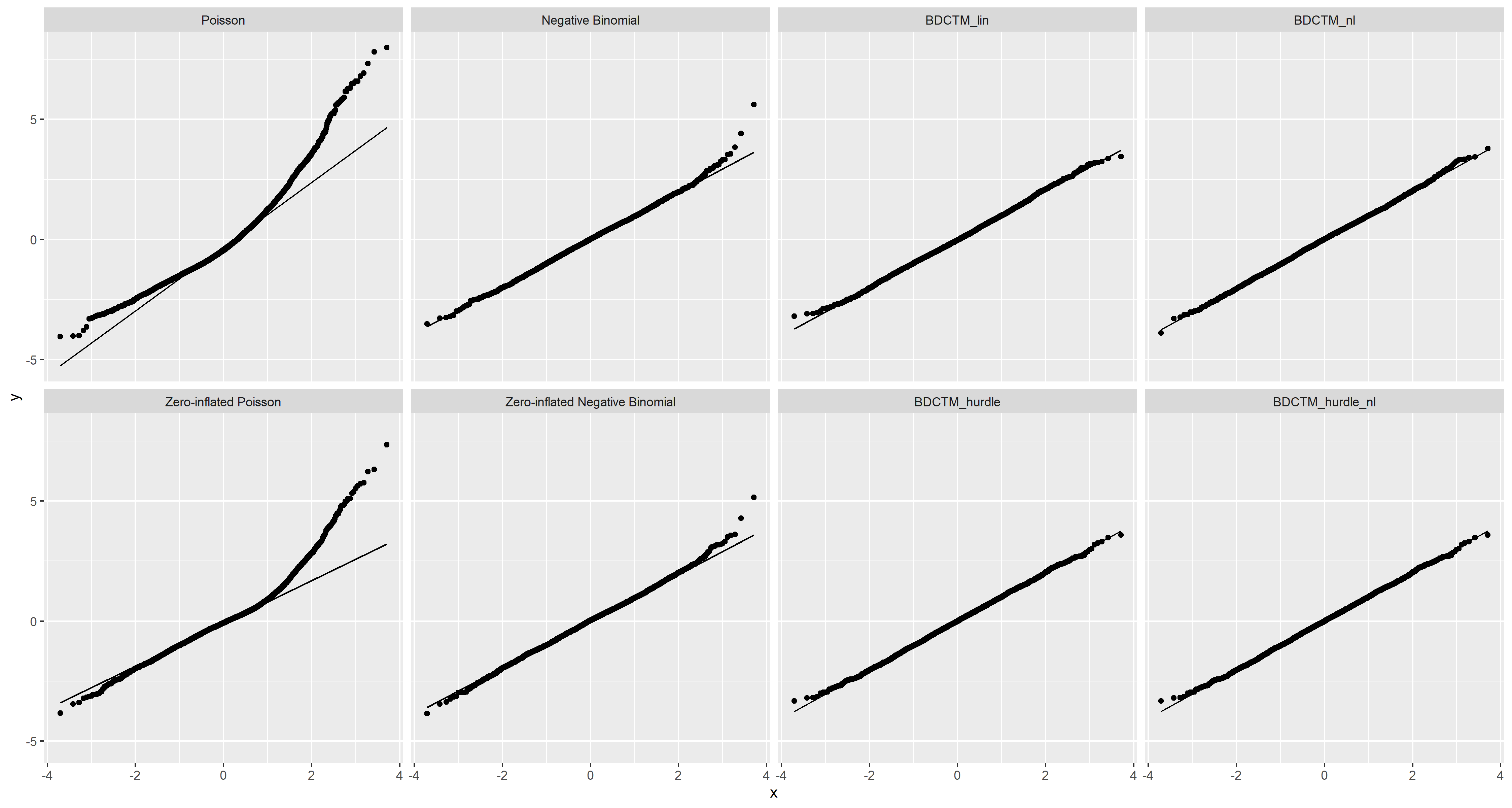}}\caption{\footnotesize Patent citations. Comparison of quantile residuals obtained by BDCTM models with and without addititional zero component with various generalized linear and zero-inflated models.}\label{fig:patent_qr}
   \end{figure}

In the next step we compared all models in terms of \textit{randomized quantile residuals} as proposed by \cite{rigby2008instructions}. For every observation $y_i$, we computed residuals $\hat{r}_i = \Phi^{-1}(u_i)$ where $\Phi^{-1}$ is the quantile function of the standard normal distribution and $u_i$ is randomly drawn from $\mathbb{U}(F(y_i-1)|\hat{\gvec}), F(y_i | \hat{\gvec}))$ with plugged in estimates $\hat{\gvec}$. $F(\cdot |\hat{\gvec})$ is the estimated conditional distribution function. Residuals obtained from the true model follow a standard normal distribution, which is why deviations can be checked by quantile-quantile plots. Fig. \ref{fig:patent_qr} shows the Q-Q plots of the considered models. Again, the Poisson model reveals a lack of fit represented by the strong deviations from the normal line which holds also true for its zero-inflated counterpart to a somewhat lesser extend. The negative binomial models provide a considerably better fit, but seem to be surpassed by the BDCTMs which indicate the best aptitude for infering the distribution of patent citations while at the same time providing a flexible "sans souci" approach, abolishing the need to search for the ``right" count distribution in general.
 
For a more rigorous assessment of the out-of-sample performance, we conclude our analysis with an evaluation based on \textit{proper scoring rules.} Originally proposed by \cite{gneiting2007strictly},  they serve as summary measures for the predictive power of a model. Based on data $y_1, \ldots, y_R$ in a validation sample and estimated probabilities $\hat{\matr{p}}_r = (\hat{p}_{r0}, \hat{p}_{r1}, \ldots)$ obtained from the predictive distribution $\hat{p}_{rk}=f(y_r =k|\hat{\gvec})$, scores are computed by taking the sum of the individual score contribution $S = \sum_{r=1}^R S(\hat{\matr{p}}_r, y_r)$.
We consider the three most prominent scores
\begin{itemize}
\item Brier score: $S(\hat{\matr{p}}_r, y_r)=-\sum_k ( \mathbbm{1}(y_r =k) - \hat{p}_{rk})^2$,
\item Logarithmic score: $S(\hat{\matr{p}}_r, y_r) = \log(\hat{p}_{ry_r})$ (out-of-sample likelihood), and
\item Spherical score: $S(\hat{\matr{p}}_r, y_r) = \frac{\hat{p}_{ry_r}}{\sqrt{\sum_k \hat{p}^2_{rk}}}$.
\end{itemize}
The probabilistic forecasts collected in $\hat{\matr{p}}_r$ for the responses $y_r$ are assessed by 10-fold cross-validation. Table \ref{tab:scores} shows the score sums obtained from the four BDCTM models introduced in this section, together with the Watanabe Information Criterion for Bayesian models (WAIC, \cite{watanabe2010asymptotic}). The \texttt{cotram} model is specified equivalently to $\text{BDCTM}_\text{lin}$, which is why their similar performance in terms of quadratic and spherical score is not surprising. Note that the logarithmic score considers only one probability of the predictive distribution and is therefore vulnerable to outliers and extreme observations which could explain the better performance of $\text{BDCTM}_\text{lin}$ in that regard. Both, considering excess zeros and nonlinear effects comes with improved predictive power culminating in the $\text{BDCTM}_{\text{hurdle-nl}}$'s dominating performance across all measures besides the WAIC where the zero component did not lead to improvements. The scores could be further improved by a model selection procedure as shown in \cite{klein2015bayesian}.
\begin{center}
\begin{table}[t]
\begin{tabular}{l c c c c}
\hline
\hline
  Model & Logarithmic & Quadratic  & Spherical & WAIC\\
 \hline
$\text{BDCTM}_\text{lin}$ & $-8119.67$ & $-3444.53$ & $2530.84$ & $6257.85$\\
  $\text{BDCTM}_\text{hurdle-lin}$ & $-8091.47$ & $-3438.87$ & $2534.39$ & $6224.634$\\  
  $\text{BDCTM}_\text{nl}$ & $-8110.94$ & $-3440.52$ & $2533.98$ & $\textbf{6040.573}$\\ 
  $\text{BDCTM}_\text{hurdle-nl}$ & $\boldsymbol{-8044.69}$ & $\boldsymbol{-3427.77}$ & $\boldsymbol{2543.44}$ & $6184.174$\\ 
     $\text{cotram}$ & $-8174.92$ & $-3443.07$ & $2531.23$ & -\\ 
   \hline
\end{tabular}\caption{Patent citations. Score sums of all models obtained via 10-fold cross-validation. Calculation of the WAICs on basis of the whole data set. Best results are depicted in bold font.}
\end{table}\label{tab:scores}
\end{center}

\subsection{A partial proportional odds model for forest health assessment}\label{sec:app2}
This short analysis involving nonlinear category-specific effects is based on data from the forest of Rothenbuch (Spessart) over the years (1982-2004). Every year, the health status is evaluated and categorized by the response variable $\mathit{defol}$ measuring defoliation grades. Since data is sparse in some of the original nine categories $(0\%, 12.5 \%, \ldots, 100\%)$, we aggregated them into the three defoliation grades $1=\text{no }(0\%), 2 = \text{weak }(12.5 \% - 37.5\%)$ and $3=\text{severe }(\geq 50 \%)$. Among others, the dataset comes with the covariates  $\mathit{canopy}$ (canopy density in percent), x, y (x- and y-coordinates of location) and $\mathit{id}$ (tree location identification number.). Check \cite{fahrmeir2007regression} for a full description of the dataset. The goal of this analysis is to determine the effect of the covariates on the degree of defoliation. Since the forest data is notorious for confounding and high autocorrelation, we let the sampler run for $10,000$ iterations with a burn-in and warm-up phase of length $1000$.

\begin{figure}
\centering  
   %\makebox[0pt]{\includegraphics[scale=0.9]
      \makebox[0pt]{\includegraphics[scale=0.5]
   {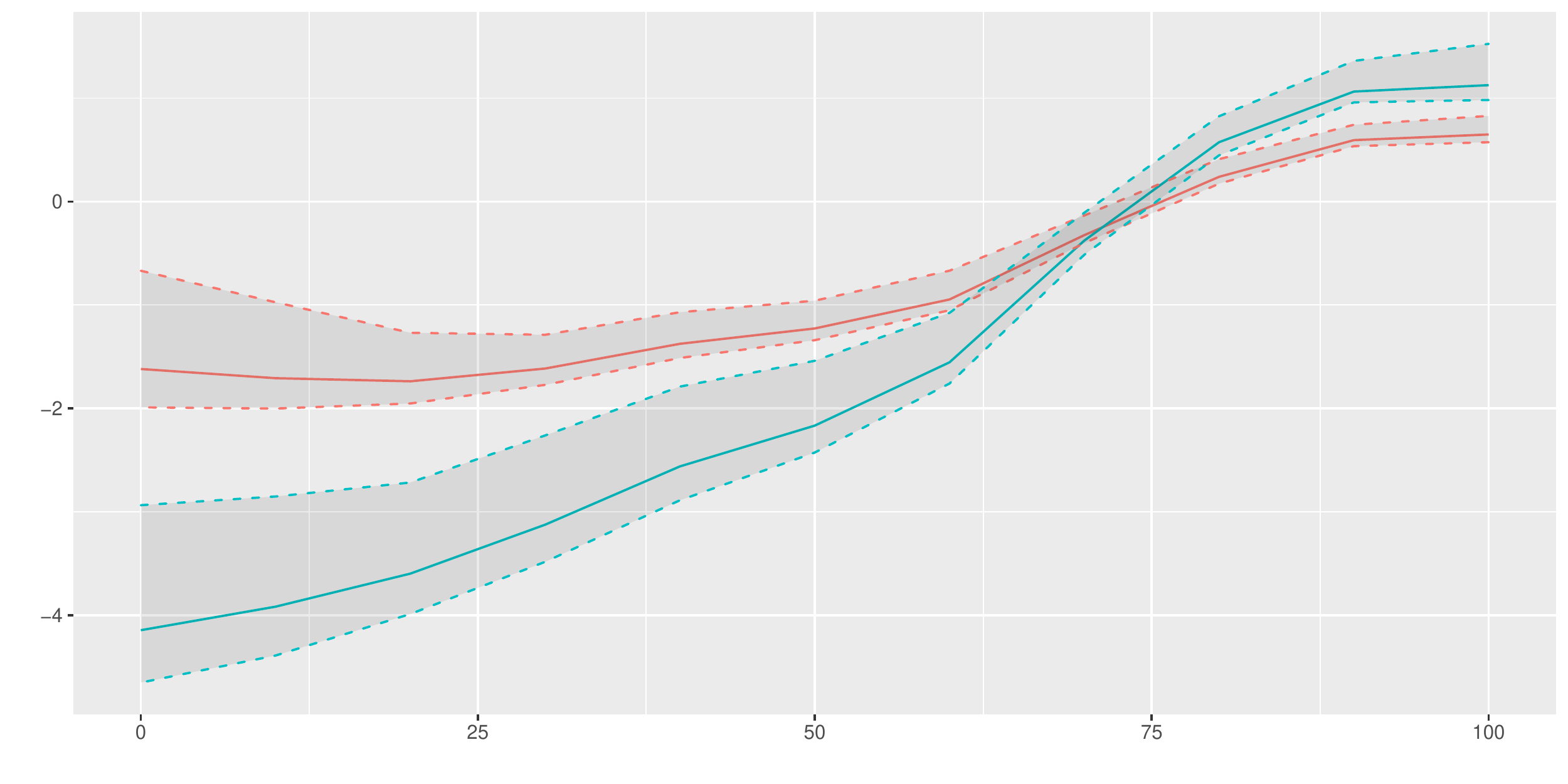}}\caption{Forest health: estimated nonlinear category-specific effect of $canopy$,  "no defoliation" in red, "severe defoliation" in blue together with $95\%$-credible intervals.}\label{fig:nl_npo}
   \end{figure}

\begin{figure}
\centering  
   %\makebox[0pt]{\includegraphics[scale=0.9]
      {\includegraphics[width=\textwidth]
   {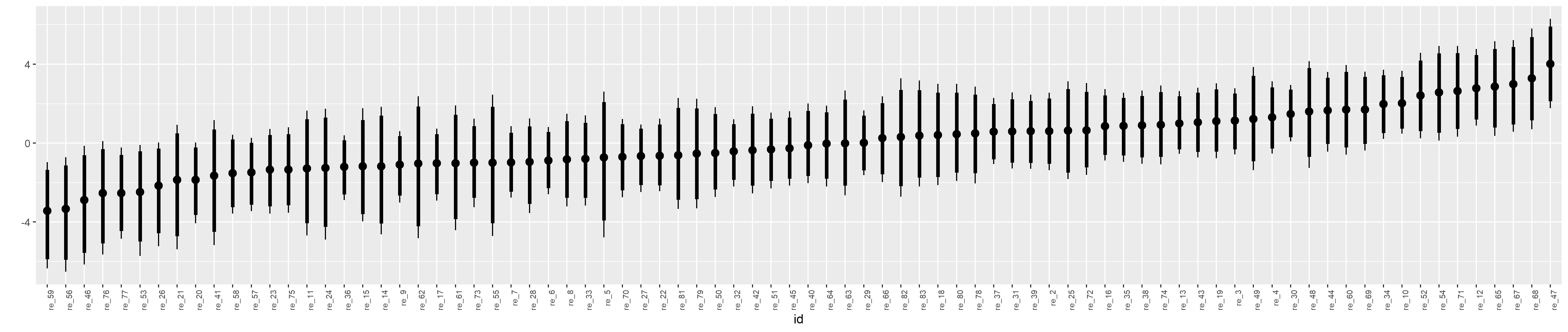}}\caption{Forest health: median-sorted estimated random intercepts for tree location groups.}\label{fig:forest_re}

   \end{figure}

\begin{figure}
\centering  
   %\makebox[0pt]{\includegraphics[scale=0.9]
      \makebox[0pt]{\includegraphics[width=\textwidth]
   {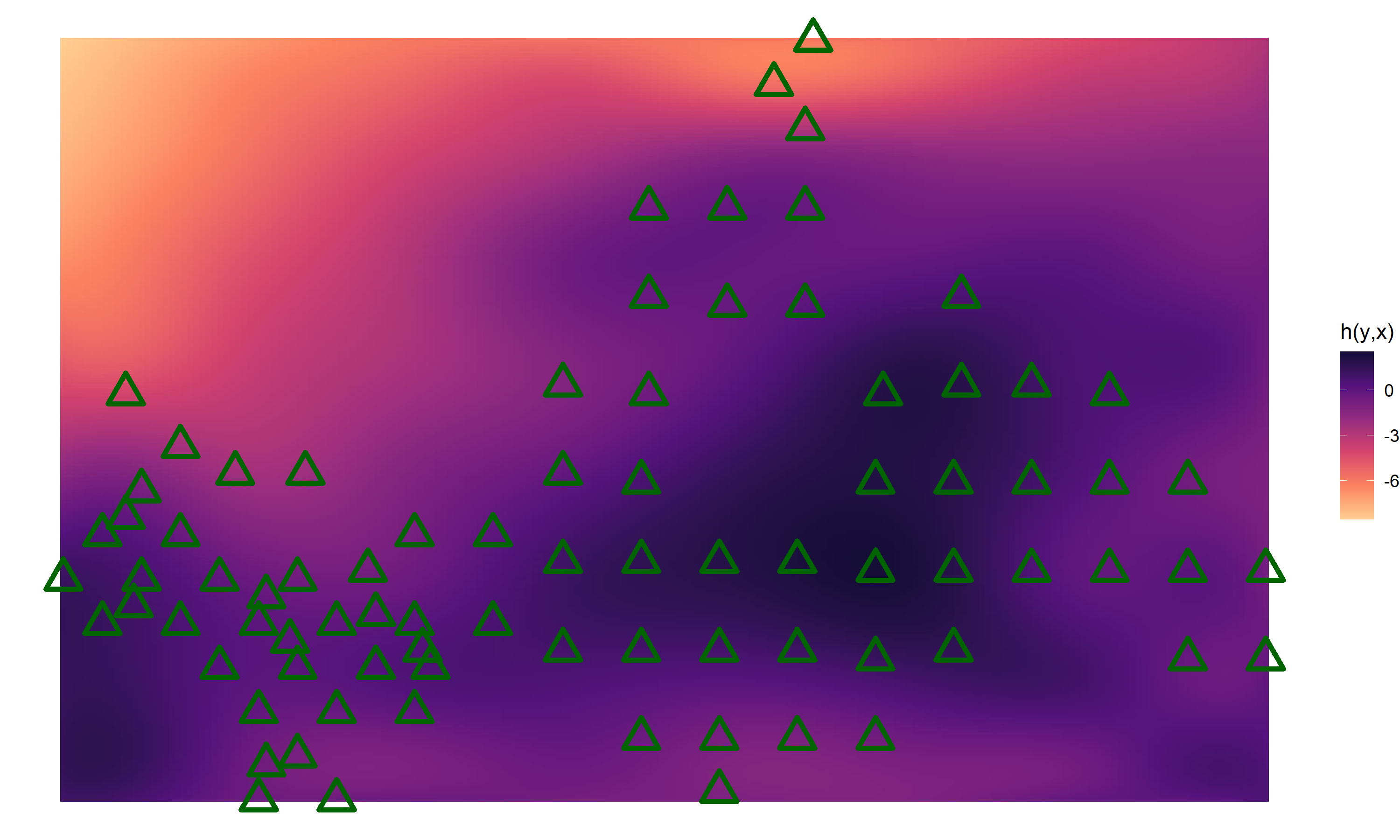}}\caption{Forest health: estimated two-dimensional spatial effect with triangles indicating observed tree locations based on 2nd order penalties.}\label{fig:forest_spat}
\end{figure}

For this, we set up the partial proportional odds model
\begin{align*}
F_{Y|\matr{X}=\matr{x}}(y_r) &= F_\text{SL}(\matr{e}(\mathit{defol})^T\gvec_1 + (\matr{e}(\mathit{defol})^T \otimes \matr{b}_{(10)}(\mathit{canopy})^T)^T \gvec_2 \\
& - \matr{b}(\mathit{id})^T\bvec_3\\
   & -(\matr{b}_{(10)}(x)^T \otimes \matr{b}_{(10)}(y)^T)^T\bvec_4),
\end{align*}
where we assume nonlinear category-specific shifts of $\mathit{canopy}$, a transformation random effect for the tree location groups and a spatial nonlinear effect on basis of a tensor spline for the coordinates $x$ and $y$. Figure \ref{fig:nl_npo} shows the estimated nonlinear category-specific effect for $\mathit{canopy}$. The section for $0 \leq \mathit{canopy}  \leq 25$ displays almost parallel curves which then vary more and more individually until they even cross. The variance of the estimated random effect for $\mathit{id}$ is $2.42$ and the standard deviation is $1.55$. Figure \ref{fig:forest_re} shows the estimated random intercepts. In a preliminary run we observed the same problems with confounding in location-specific effects as \cite{fahrmeir2007regression} which could be improved to some extend by adding the spatial effect. It is displayed in Figure \ref{fig:forest_spat}. 

\section{Discussion}\label{sec:end}
With the Bayesian discrete conditional transformation model we present a novel Bayesian model framework for discrete data that combines cumulative link models with models for count data through directly modeling the conditional distribution function. Approaching these discrete data structures from the transformation perspective allows us to unify models that are usually treated seperately under the same umbrella. 
The BDCTM is flexible in the sense that it permits the user to control interpretability by means of choosing a reference distribution in conjunction with an additive transformation function. Estimating the conditional distribution function directly makes deriving distributional aspects such as the conditional quantiles straightforward by numerical inversion of $F_Z(h(y|\matr{x})$ \citep{ctm}. Furthermore, our Bayesian inferential procedure lets us obtain credible intervals and other quantities of interest without having to rely on large sample approximations. All high-dimensional effects are joined with suitable prior specifications resulting in smooth effects across the board.

We demonstrate BDCTM's ability to handle under- or overdispersion in an adaptive fashion without restrictive distributional assumptions in Sections \ref{sec:sims} and \ref{sec:apps}. A short investigation of a nonlinear non-proportional odds model highlights the versatility of our approach. In a model selection context, the unifying scope of the transformation function turns out to be a valuable simplification because there is just one ``predictor" that has to be constructed. Though not shown in this article, it is possible to establish a relationship between overdispersion and the covariate effects by including full nonlinear interactions between the count response and the respective explanatory variable. Constructing the conditional transformation function can be difficult as informed decisions about which effects to include and to interact with the response are required. Therefore, it would be desirable to develop an effect selection strategy via spike and slab priors in the spirit of \cite{klein2021bayesian} for the BDCTM that could effectively tell the user what kind of effect is impacting the regular count process, the zero component or overdispersion.

As demonstrated in Section \ref{sec:app2}, our cumulative link transformation approach can be supplemented with category-specific linear or nonlinear effects by modeling them as response-covariate interactions. This way, popular models such as (non-)proportional odds or hazards models can be retrieved simply by specifying the reference distribution. 
Both the count and the ordinal model could be supplemented with a more flexible link function as proposed by \cite{aranda1983extension}, i.e.
\begin{align*}
F(h) &= 1-(\lambda \exp(h) + 1)^{-\lambda^{-1}},
\end{align*}
which depends on an auxiliary parameter $\lambda \in ]0, \infty [$, mitigating between the log-log link for $\lambda \rightarrow 0$ and the logistic link when $\lambda \rightarrow 1$.  \cite{horowitz2001nonparametric} avoided specifying the link function entirely. A Bayesian version would entail prior distributions on the space of nonparametric continuous reference distribution.

To conclude, we believe that in this article, the BDCTM is established as a flexible, modular modeling framework in the world of discrete data that is competitive in many modern scenarios.

%%% Acknowledgements (if any)
%%% ------------------------------------------
\section*{Acknowledgements}
The work of Manuel Carlan was supported by DFG via the research training group 1644. Thomas Kneib received financial support from the DFG within the research project KN 922/9-1.

%%% References if bibTeX is used
%%%
%%% Please, do not specify any \bibliographystyle{} command!
%%%
%%% It is already specified in the smj.cls and its
%%% second specification here causes error.
%%% ------------------------------------------------------------

\bibliography{lit}

\end{document}